\documentclass[journal=nalefd,manuscript=letter,layout=twocolumn]{achemso}

\usepackage{graphicx}
\usepackage{dcolumn}
\usepackage{bm}
\usepackage{gensymb}

\usepackage{hyperref}
\hypersetup{
    colorlinks=true,
    linkcolor=blue,
    filecolor=magenta,      
    urlcolor=cyan,
    }

\usepackage[utf8]{inputenc}
\usepackage[english]{babel}
\usepackage{amsmath}
\usepackage{graphicx}
\usepackage{xcolor}
\usepackage{float}
\usepackage[version=4]{mhchem} 
\usepackage{siunitx}
\usepackage{tabularx}
\usepackage{dcolumn}
\usepackage{nameref} 

\title{Novel Electrical Characterization Method for Antiferroelectrics using a Positive Up Negative Down Approach}

\author{Grégoire Magagnin}
\affiliation[]{Ecole Centrale de Lyon, INSA Lyon, CNRS, Universite Claude Bernard Lyon 1, CPE Lyon, INL, UMR5270, 69130 Ecully, France}
\email{gregoire.magagnin@ec-lyon.fr}

\author{Martine Le Berre}
\affiliation[]{INSA Lyon, Ecole Centrale de Lyon, CNRS, Universite Claude Bernard Lyon 1, CPE Lyon, INL, UMR5270, 69621 Villeurbanne, France}

\author{Sara Gonzalez}
\affiliation[]{CNRS, INSA Lyon, Ecole Centrale de Lyon, Universite Claude Bernard Lyon 1, CPE Lyon, INL, UMR5270, INSA, 69621 Villeurbanne Cedex, France}

\author{Damien Deleruyelle}
\affiliation[]{INSA Lyon, Ecole Centrale de Lyon, CNRS, Universite Claude Bernard Lyon 1, CPE Lyon, INL, UMR5270, 69621 Villeurbanne, France}

\author{Bertrand Vilquin}
\affiliation[]{Ecole Centrale de Lyon, INSA Lyon, CNRS, Universite Claude Bernard Lyon 1, CPE Lyon, INL, UMR5270, 69130 Ecully, France}

\author{Jordan Bouaziz}
\affiliation[]{Ecole Centrale de Lyon, INSA Lyon, CNRS, Universite Claude Bernard Lyon 1, CPE Lyon, INL, UMR5270, 69130 Ecully, France}
\email{jordan.bouaziz@ec-lyon.fr}


\begin{document}

\begin{abstract}
    This study demonstrates the effectiveness of AFE-PUND, a revisited Positive Up Negative Down (PUND) protocol for characterizing antiferroelectric (AFE) materials, in analyzing $ZrO_2$ films across different thicknesses, revealing key trends. The proposed AFE-PUND method enables the isolation of switching currents from non-switching contributions, allowing precise extraction of remanent polarization and coercive field from hysteresis loops. The remanent polarization increases with film thickness, reflecting enhanced domain stability, while endurance cycles highlight the wake-up effect and its eventual degradation due to fatigue in thicker films. Similarly, coercive fields decrease with thickness, indicating reduced switching barriers and a clearer transition between tetragonal and orthorhombic phases. The method provides valuable insights into micro-structural influences, such as defect accumulation, grain size, and domain wall pinning, which critically affect device performance. AFE-PUND thus establishes itself as an essential tool for advancing the understanding and optimization of antiferroelectric materials.
\end{abstract}

\maketitle

\section{Introduction}

\indent
    The study of ferroelectric (FE) and antiferroelectric (AFE) materials has gained significant attention due to their promising applications in next-generation electronic devices, such as non-volatile memories, energy storage, and neuromorphic systems \cite{Dawber2005,christensen20222022,hao2013review}. Central to advancing these technologies is the ability to accurately characterize the electrical behavior of these materials under varying conditions. One critical method for such characterization is the Positive Up Negative Down (PUND) protocol, originally devised for FE materials, which enables the isolation of switching current contributions from other transient electrical components. The present work revisits and adapts this method for AFE materials, offering a novel approach for distinguishing their unique switching dynamics.

\indent
    The concept of the PUND method was first introduced by Scott and al. \cite{scott1988switching} for ferroelectrics (FEs) in 1988. In this initial article, authors explained how, when starting from the negative spontaneous polarization, $-P_s$, if a positive switching voltage is applied, the electric induction $\vec D$ (also called electric flux density or more frequently: electric displacement field) is: $D = 2P_s+\varepsilon_0 \varepsilon_r E$ (along z direction in the case of a simple planar capacitor), where $\varepsilon_r$ is the dielectric constant of the material, $\varepsilon_0$ the dielectric constant of vacuum, and E is the applied electric field. Authors noticed that the thin film they studied ($PbZr_{0.54}Ti_{0.46}O_3$) is then already polarized positively and if a second positive (non-switching) pulse is applied, $D = \varepsilon_0 \varepsilon_r E$. Consequently, subtracting the first and second values of D allows the extraction of the film's switching polarization. In this article, the term "PUND" is not yet mentioned. The first occurrence of "PUND" might date back to 1989 in an article by Scott et al.\cite{scott1989hb}.  
\\
\indent 
    The PUND method is widely utilized in standard tests for capacitors, including endurance tests, which consist in measuring the number of cycles that can be applied before breakdown. During testing, current response to voltage pulses for FE materials can increase or decrease depending on the number of cycles. The degradation of ferroelectric properties upon cycling, notably the decrease in switching current ($I_{switch}$), is called fatigue. On the contrary, wake-up (WU) effect consists in an increase of $I_{switch}$ during the early stages of the endurance tests. The PUND method allows to study more in detail physical mechanisms such as fatigue and WU effect. By isolating $I_{switch}$ from the dielectric and leakage contributions (respectively $I_\varepsilon$ and $I_L$), it further helps to understand which contribution is playing a critical role. Quite often current response also increases just before breakdown. Thanks to the PUND method, the WU effect can be easily differentiated from the increase in leakage currents, for example. More specifically, in doped-$HfO_2$ it has been demonstrated that the WU effect is not due to an increase of leakage, but to an $I_{switch}$ increase due to material reorganization (oxygen vacancies reorganization and/or electrons trapping/de-trapping, transition of domains from one phase to another, domains pinning/depinning, etc.). \cite{lee2023role, mcmitchell2021elucidating, jiang2021wake, lederer2021origin} Finally, PUND method also avoids any problem of the "BANANA effect"\cite{scott2007ferroelectrics}.
\\
\indent 
    For antiferroelectrics (AFEs), however, the application of PUND presents unique challenges. Unlike FEs, AFEs exhibit two switching current peaks under the application of a voltage pulse: one during voltage rise and the other during the voltage fall. When the voltage goes back to zero, the material has already switched positively and negatively, and the polarization goes back to zero, leading to: $I(P)-I(U)=I(N)-(D)=0$. Thus, the standard PUND protocol must be adapted.
\\
\indent 
    This article proposes an automatized protocol in order to use the PUND principle for AFE materials. To highlight the method, planar Metal/Insulator/Metal (MIM) capacitors with $ZrO_2$ as AFE insulator are employed. The procedure can, of course, be applied to other AFE materials as well.

\section{Method} \label{sec:method}

\begin{figure*}[!ht]
    \centering
    \includegraphics[width=\textwidth]{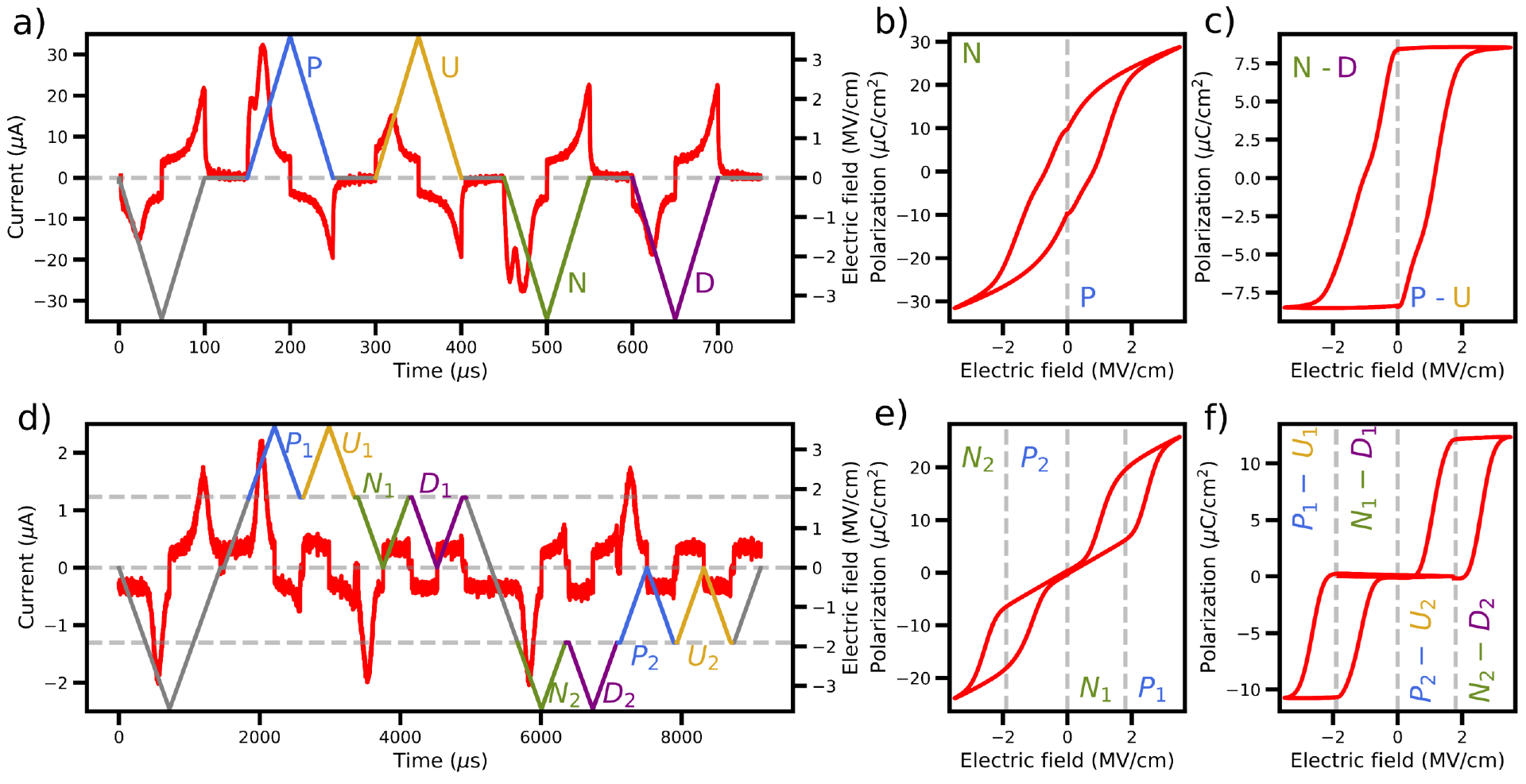}
    \caption{Principle of the PUND method for FE and AFE materials. Panel (a) illustrates the PUND sequence, with the electric field (right axis) applied to FE materials and the corresponding current response (red curve, left axis) as a function of time. This sequence isolates the FE contribution. Panel (b) shows the polarization-electric field (P-E) hysteresis corresponding to the "PN" sequence, while panel (c) presents the P-E hysteresis derived from the "P-U" and "N-D" sequences, with dielectric and leakage contributions removed to isolate the FE response. Panel (d) displays the PUND sequence for AFE materials, showing the electric field (right axis) and the corresponding current response (red curve, left axis) over time. This sequence isolates the AFE contribution by distinguishing the two FE contributions. Panel (e) provides the P-E hysteresis corresponding to the "PN" sequences for AFE materials, and panel (f) shows the P-E hysteresis derived from the "P-U" and "N-D" sequences, isolating the AFE response after removing dielectric and leakage effects.}
    \label{fig:1}
\end{figure*}

\indent
    This section begins with a summary of the PUND method for FEs, before explaining the approach developed for AFEs. In the following, the technique dedicated to AFEs will be called AFE-PUND to make a difference with the process for FEs, subsequently designated as FE-PUND.
\\
\indent 
    Figure \ref{fig:1}(a) is reminding the FE-PUND method principle. The pulse train for the acquisition of the P-V or P-E characteristics is the PUND schema, which consists on a train of five triangular pulses with the same amplitude ($\pm V_{max}$) and rise and fall times ($t_{rise}$, $t_{fall}$). Each pulses being separated by a relaxation time ($t_{relax}$) where no electrical field is applied. For greater clarity, $V_{max}$ and these characteristic times are explicitly shown in the supplemental (figure S1). The method can be described as follows:
\begin{itemize}
    \item \textbf{Pre-switch pulse}: this pulse aligns the ferroelectric domains along one specific direction.
    \item \textbf{First positive pulse (P)}: to switch ferroelectric domains upward, it includes contributions due to domain motion, dielectric and leakage currents.
    \item \textbf{Second positive pulse (U)}: to measure other contributions to the current, which are not due to the ferroelectric domain switching.
    \item \textbf{First negative pulse (N)}: it switches domains downward, it includes other contributions due to the domains motion, dielectric and leakage currents in the opposite sense.
    \item \textbf{Second negative pulse (D)}: as for the U pulse, it is used to measure other contributions to the current, which are not due to the domain switching.
\end{itemize}
\indent 
    As the same reasoning applies when subtracting the P and U pulses or the N and D pulses, for this example, a focus is made on P and U. The hysteresis loop is obtained by subtracting the response of the U pulse from that of the P pulse, which allows for the isolation of the specific contributions of each pulse. 
\\
\indent
    By definition, the current is the infinitesimal change in charge carriers dq with respect to the differential of time dt. Considering an electrical excitation $\varepsilon_0 \vec E$ (an applied electrical field in our case), a change in $\vec D$ is induced. The charge can then be rewritten as the scalar product of $\vec D$ and the surface\nobreakspace{vector $\vec S$}.
    \begin{equation}
        I = \frac{dq}{dt} = \frac{d \left( \vec D \cdot \vec S \right)}{dt} 
    \end{equation}

For a planar capacitor, considering that the PUND pulse train is applied, the equation can be rewritten along the z axis as follows: 
\begin{equation}
    D = \frac{1}{S} \int_0^{t_{rise}+t_{fall}} Idt
    \label{eq:D(I)}
\end{equation}
\indent
    Figure \ref{fig:1}(b) represents the polarization versus voltage (P-E) curve derived from the current under the P and N peaks according to equation \ref{eq:D(I)}. From a technical standpoint, this hysteresis curve should be referred to as D-E curves rather than true P-E curves. It is also the case of typical pulse sequences used for Dynamic Hysteresis Measurement (DHM) in literature. On the contrary, the PUND technic allows to plot true P-E curves as it will be now demonstrated. 
\\
\indent
    It is assumed here that the pulse takes a triangular shape. An additional time value should be added if pulses have trapezoidal shapes. Here, $t_{rise} = t_{fall}$ leading to $t_{rise}+t_{fall} = 2 t_{rise}$. As there is no electrical field applied during $t_{relax}$, it can be considered that during P and U pulses the equation to isolate the switching transient current can be expressed as follows:

\begin{flalign}
    & D_{PU} = \frac{1}{S} \int_{0}^{2t_{rise}} I(P) - I(U) dt \\
    & D_ {PU} = \frac{1}{S} \int_{t_P}^{t_P+2t_{rise}} I(P) dt - \frac{1}{S} \int_{t_U}^{t_U+2t_{rise}} I(U) dt \\ 
    & D_{PU} = D(P) - D(U) 
\end{flalign}

D(P) and D(U) are here defined. 
\\
If now the expression of the electric induction is rewritten as a function of the applied electric field and the polarization, it is very well known that: 
\begin{equation}
    \vec D = \varepsilon_0 \varepsilon_r \vec E = \varepsilon_0  \vec E + \vec P
\end{equation}

For a planar capacitor, along z axis, using equation (5) and (6) it gives:
\begin{equation}
    D_{PU}  = \varepsilon_0  E + P(P) - \varepsilon_0  E - P(U)
\end{equation}

\begin{figure}[!ht]
    \centering
    \includegraphics[width=0.47\textwidth]{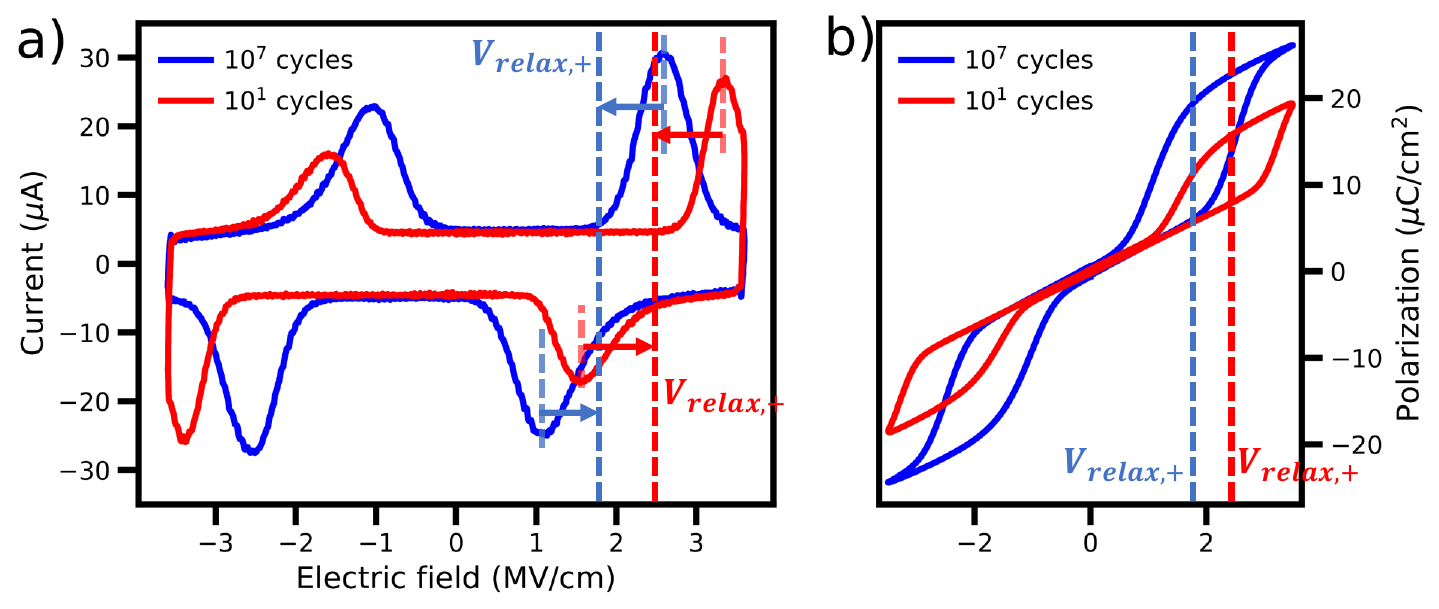}
    \caption{Necessity to follow the $V_{{relax},{+}}$ and $V_{{relax},{-}}$ biases with cycling. (a) Current response over the applied voltage for a \SI{10}{nm}-thick $ZrO_2$ after $10$ and $10^7$ cycles. The values of $V_{{relax},{+}}$ and $V_{{relax},{-}}$ are determined by extracting the displacement current peak positions at each cycle. (b) Corresponding P-E curves obtained by the DHM method with dashed lines representing the hysteresis the extracted $V_{{relax},{+}}$ biases for two cycle states.}
    \label{fig:2}
\end{figure}

And finally the ferroelectric contribution to polarization can be isolated by the equation:

\begin{equation}
    D_{PU} = P_{PU}  = P(P) - P(U)
    \label{eq:P_PU}
\end{equation}
\indent
    The polarization values in Figure \ref{fig:1}(c) are obtained using equation \ref{eq:P_PU} for the current under P-U, and similarly for N-D. This demonstrates that P has the dimension of a polarization, as previously stated, contrary to DHM and PN curves where it has the dimension of an electric induction.
\\
\indent 
    The voltage during the relaxation time (equal to zero in Figure \ref{fig:1}(a)) is called the "relaxation voltage" (sometimes also called "resting voltage"). The relaxation time is used to highlight back-switching (BS). As its name implies, BS consists in the FE domains switching back in the opposite direction of polarization in which they were, before reaching the relaxation voltage. During $t_{relax}$, particularly in the case of asymmetrical ferroelectric structures, FE domains can start to switch in the opposite polarization direction before reaching, in the example figure \ref{fig:1}(a), zero volt or at zero volt. This cause the presence of domains still switching during the application of U or D voltage pulses. This BS can be clearly seen figure \ref{fig:1}(a) under each voltage pulses and will be re-discussed in section \nameref{sec:limitations}. 
\\
\indent 
    Finally, details regarding capacitor fabrication have already been provided in our earlier work\cite{magagnin2024comparative}. And FE material selected to illustrate the PUND method here is $Hf_{0.5}Zr_{0.5}O_2$ (HZO). The HZO sample used in this article does not represent the highest performance achievable; capacitors with superior characteristics (high $P_r$ and without BS) were presented in our prior studies \cite{magagnin2024comparative,bouaziz2019huge}. Nevertheless, this sample was specifically chosen to emphasize the BS mechanism and later demonstrate that BS can also occur for AFE-PUND.  
\\
\indent
    To conclude, by subtracting the current under the P and U peaks, integrating the result, and normalizing by the surface area, the polarization due to ferroelectric switching can be expressed as a function of the applied electric field, with all other contributions effectively removed. Figure \ref{fig:1}(b) shows a hysteresis curve that includes all contributions to the current, while the outcome of this subtraction is presented in Figure \ref{fig:1}(c).
\\
\\
\indent
    The challenge with AFE materials lies in the fact that each pulse induces two switching current peaks. To isolate these peaks, the relaxation voltage must be set to a nonzero value. An AFE material can be schematically seen as a combination of a dielectric material and two ferroelectric materials, one switching at positive coercive voltages $V_{c,+,up}$ and $V_{c,+,down}$ and the other at negative voltages $V_{c,-,up}$ and $V_{c,-,down}$. Consequently, by employing two PUND sequences—one with a positive relaxation voltage and one with a negative relaxation voltage-the same logic used for ferroelectrics can be extended to antiferroelectrics. The pulse train for the two PUND sequences with positive and negative relaxation voltages is shown in figure \ref{fig:1}(d). The corresponding P-E curve is presented in \ref{fig:1}(e) for the current under $P_1$, $N_1$, $P_2$, and $N_2$ voltage peaks, while the subtraction of the different contributions is illustrated in \ref{fig:1}(f). Between two pulses of the first $P_1U_1N_1D_1$ or second $P_2U_2N_2D_2$ sequence, there is a relaxation period during which the electrical field remains at the relaxation electrical field. Therefore, the relaxation time for AFE-PUND is nonzero. However, in figure \ref{fig:1}(d), the relaxation time has been set to \SI{50}{\micro \second}, which is too short to be visually discernible in figure \ref{fig:1}(d).
\\
\indent
    The sole remaining challenge at this stage is to determine the appropriate voltage values for the relaxation voltages. In figure \ref{fig:1}(d), an automatic detection has been used prior to AFE-PUND sequence. On Figure \ref{fig:2}, this automatic detection procedure of the relaxation voltages is explained. 

\indent 
    From a mathematical point of view, the coercive electrical field ($E_c$) corresponds to the maximum of the slope of the P-E curve, so the maximum of $dP/dE$. In theory, this maximum is infinite. It can be mathematically demonstrated from equation (3) and $\vec E = - \overrightarrow{grad} ~ V$ that the coercive voltage will therefore corresponds to the position of the maximum $I_{switch}$ of a FE material (if the FE material is not leaky). In the I-V curve (Fig\ref{fig:2}(a)), the position of the maxima on the y-axis (current axis) corresponds to the coercive voltage ($V_c$) on the x-axis (voltage axis). It can be determined four coercive voltages for an AFE material : $V_{c,+,up}$, $V_{c,+,down}$, $V_{c,-,up}$, $V_{c,-,down}$.
    The detection of the relaxation voltage is then determined by the equations : 
    \begin{equation}
        V_{relax,+} = \frac{|V_{c,+,up}|+|V_{c,+,down}|}{2}
    \end{equation}
    \begin{equation}
        V_{relax,-} = \frac{|V_{c,-,up}|+|V_{c,-,down}|}{2}
    \end{equation}
    Once the relaxation positive and negative voltages are determined, the two PUND sequences are automatically created via a Python code. 
\\
\indent
   In conclusion, by using a DHM sequence for the automatic detection of the relaxation voltages, it is possible to create automatically a pulse train made of two PUND sequences using a positive and a negative relaxation point (respectively $V_{relax,+}$ and $V_{relax,-}$) and to create a AFE-PUND sequence that allows to create curves that keeps only the ferroelectric positive and negative contributions of antiferroelectric materials, removing dielectric and leakage non-switching transient components of current.

\section{Results and Discussion}

\begin{figure*}[!ht]
    \centering
    \includegraphics[width=\textwidth]{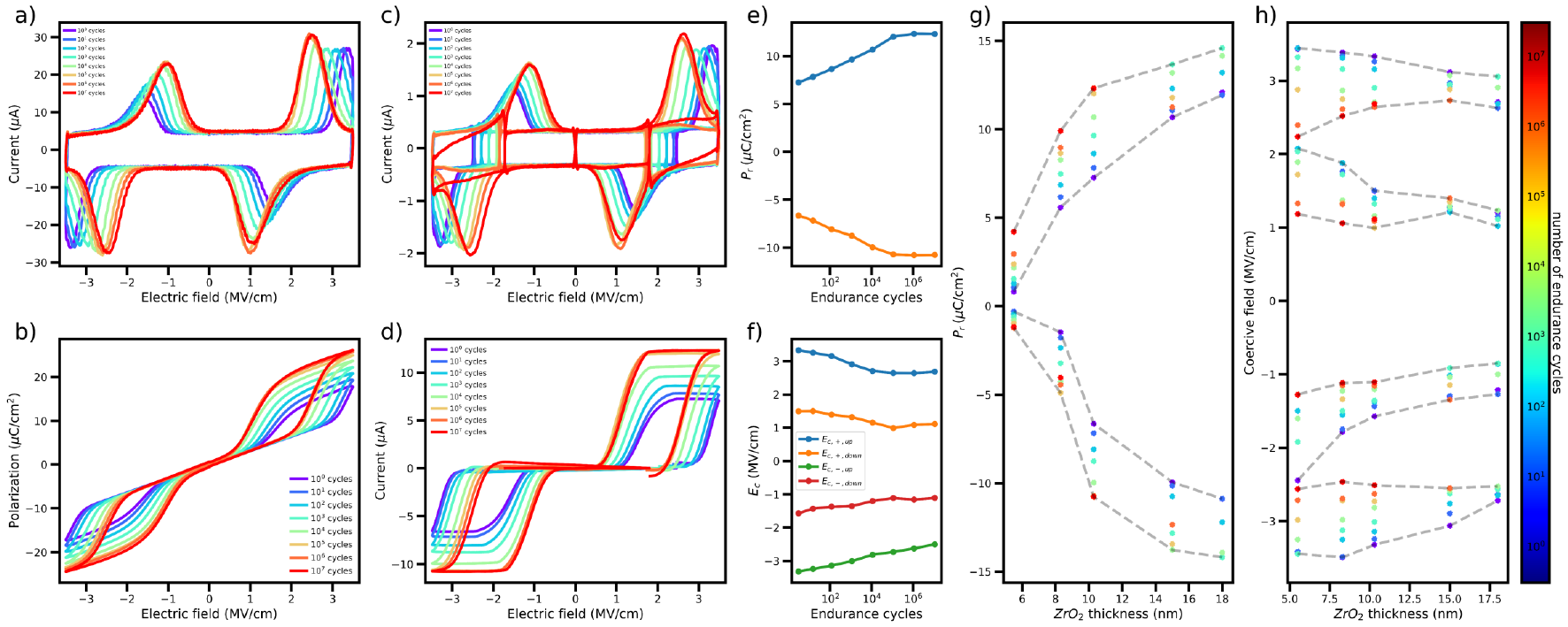}
    \caption{(a) to (f) Electrical characterization of a AFE capacitor with $ZrO_2$ (10 nm) at 3.5 MV/cm. DHM protocol: corresponding (a) current versus voltage as a function of cycle number and (b) P-E loops as a function of cycle number. AFE-PUND protocol: corresponding (c) current versus voltage as a function of cycle number and (d) P-E loops as a function of cycle number. Extracted by AFE-PUND: (e) remanent up and down polarizations as a function of cycle number. (f) Coercive fields as a function of cycle number. (g) and (f) AFE-PUND protocol over several $ZrO_2$ AFE capacitors thicknesses between 5.5 nm and 18 nm at 3.5 MV/cm: g) Variation of the remanent polarization and h) variation of the coercive fields for several $ZrO_2$ thickness and endurance cycles.}
    \label{fig:3}
\end{figure*}

\indent 
    On figure \ref{fig:3}, for the comparison with our AFE-PUND sequence, a DHM sequence was used. Corresponding I-V curves (\ref{fig:3}(a)) and P-E loops (\ref{fig:3}(b)) are here shown in order to explain standard electrical characterizations of antiferroelectric materials. The DHM sequence is also used for automatic detection as explained in the section \nameref{sec:method}. This sequence has also been widely employed in prior studies characterizing antiferroelectric materials \cite{nguyen2018electric}. For the AFE-PUND sequence the I-V curves and P-E loops are shown respectively in figure \ref{fig:3}(c) and \ref{fig:3}(d). One can notice that the P-E curves show two hysteresis "flat" curves as described in previous section \nameref{sec:method}. From the P-E curves of figure \ref{fig:3}(d), saturation  polarization (here equal to remanent polarization $P_r$) and coercive fields $E_c$ can be extracted (respectively Figures \ref{fig:3}(e) and \ref{fig:3}(f)).
\\
\indent 
    In FE HZO, the WU effect consists in the increase of $P_r$ \cite{zhou2013wake} and in an $E_{c}$ shift \cite{cai2022investigation}. These two effects are observed for our $ZrO_2$ sample with 10 nm-thickness, on figure \ref{fig:3}(e) and \ref{fig:3}(f). It demonstrates that the WU effect is not the result of a leakage increase but a real change in the contribution to ferroelectricity. 
\\
\indent
    Another important contribution of the AFE-PUND method is the study of the WU effect for different thicknesses of $ZrO_2$. As shown in Figure \ref{fig:3}(g), the remanent polarization $P_r$ increases with film thickness. This behavior indicates that thicker films provide more stable domain configurations, which enhance their ferroelectric properties. Moreover, the evolution with the endurance is clearly illustrated in the color-coded points of Figure \ref{fig:3}(g). As the cycles progress, the $P_r$ initially increases for each $ZrO_2$ thickness, reflecting the WU effect. This increase is attributed to defect reorganization, such as oxygen vacancy redistribution, and the stabilization of ferroelectric domains. However, in thicker films (above 15 nm), we also observe a noticeable decrease in $P_r$ after $10^3$ cycles, indicating the onset of fatigue. This behavior arises due to defect accumulation and domain wall pinning, which hinder further polarization switching.
    
\indent    
    In addition, Figure \ref{fig:3}(h) demonstrates that the coercive field $E_c$ decreases with increasing thickness, suggesting that thicker films require a lower energy barrier for polarization switching. The evolution of $E_c$ reflects a difference in the structural phase transition from the tetragonal (linear dielectric) phase to the orthorhombic (ferroelectric) phase under the applied electric field with thickness\cite{hoffmann2022antiferroelectric}. The transition from the tetragonal to the orthorhombic phase shows a clear reduction in $E_c$ whereas the reverse transition exhibits also a reduction, but with a different dynamic. Moreover, we can also observe that $E_c$ decreases during the early endurance cycles for each thicknesses, consistent with the structural reorganization that lowers the energy barrier for domain switching as the material transitions more and more from the tetragonal to orthorhombic phase. Same as for the $P_r$ variation, we observe a different behavior for thicker films (above 15 nm): the coercive fields stabilizes after a certain number of cycles and may even exhibit signs of slight increase back, reflecting the impact of fatigue on the switching dynamics. This seems to be mainly visible on the coercive fields switching from the tetragonal to the orthorhombic phase. These endurance-dependent trends are particularly evident in the color gradients of the subplots, which highlight the contrasting behaviors between thinner and thicker films.
    
\indent
    The different FE behavior observed in $ZrO_2$ films above a certain thickness can be attributed to a combination of microstructural and electrical factors. Thicker films tend to have larger grains or several layers of grains, which increase domain wall pinning and reduce mobility, leading to fatigue and a decline in remanent polarization with endurance. Additionally, higher defect density and less uniform electric field distribution in thicker films may hinder complete polarization switching, contributing to the stabilization or slight increase of coercive fields with cycling. Stress and strain effects, potentially more pronounced in thicker films, could also alter the energy landscape for domain switching, further exacerbating fatigue. Conversely, thinner films, with smaller grains and more uniform defect profiles, exhibit enhanced WU effects due to easier domain reorganization. 

\indent    
    Removing dielectric and leakage contributions via this method allows the study of the variation in electrical parameters with higher accuracy. For instance, the large increase of $P_s$ for low $ZrO_2$ thickness, as seen in figure \ref{fig:3}(b) could be solely impeded to a $P_r$ increase, however once AFE-PUND method is employed it is demonstrated that the change in $P_r$ from pristine to woken-up sample is quite low, as shown on figure \ref{fig:3}(g). Then some part of the seen increase in $P_s$ can be attributed to a leakage increase. By providing a more accurate isolation of ferroelectric contributions, the AFE-PUND method enables a deeper understanding of the influence of film thickness and endurance on the ferroelectric behavior of $ZrO_2$, compared to the usual standard DHM. 

\section{Applicability and Limitations of the Method} \label{sec:limitations}

\indent
    These findings highlight the robustness of the AFE-PUND method in isolating ferroelectric contributions and accurately tracking the evolution of remanent polarization and coercive fields with endurance. However, challenges arise when dealing with samples exhibiting high leakage currents and back-switching phenomena. Understanding and addressing these limitations is critical to ensure the reliability of the AFE-PUND method, particularly for samples with significant leakage. This issue is explored in the next section, which discusses the impact of leakage currents.

\begin{figure}[H]
    \centering
    \includegraphics[width=0.47\textwidth]{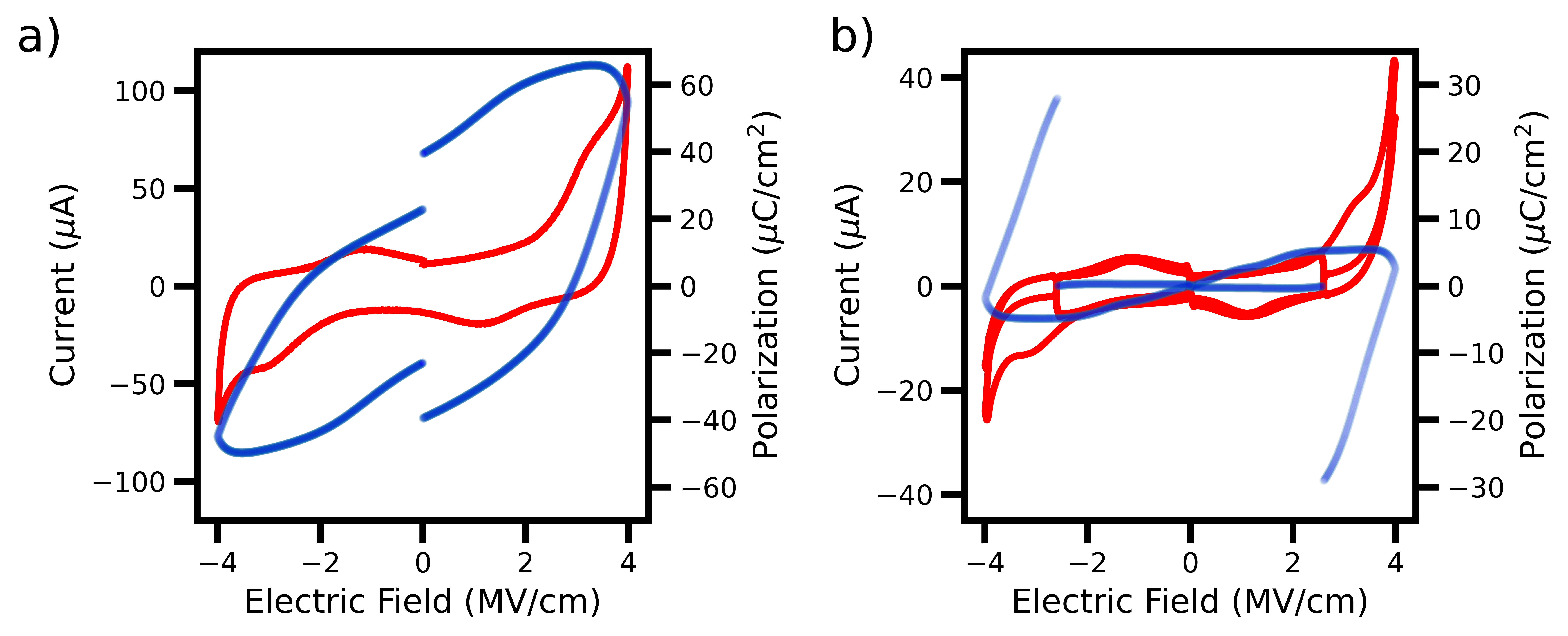}
    \caption{Leakage limitations of the AFE-PUND method: example of the AFE capacitor $TiN$/$ZrO_2$/$TiN$ with 18 nm-thick $ZrO_2$ at 4 MV/cm. (a) I-V curve and P-E for the DHM measurement; (b) I-V curve and AFE-PUND for the AFE-PUND measurement.}
    \label{fig:4}
\end{figure}

\indent
    The automatic detection of $V_{relax}$ depends on the switching peak maximum detection. When samples present high leakage, the switching peak can become less defined and the maximum less intuitive to detect reliably. Indeed, leakage increases with voltage increase in insulator dielectric oxides \cite{sze2007physics}. This is illustrated by figure \ref{fig:4}. On figure \ref{fig:4}(a) the DHM measurement is shown. Leakage current is bigger than switching current and both contributions are overlapping. Not only it leads to a loss of accuracy in the automatic detection $V_{relax}$, but also the subtractions between P and U or N and D peaks is biased. These limitations also exists with FE-PUND sequence: when the switching current is entirely overlapping with leakage current, the distinction is not possible. On \ref{fig:4}(b), the result of the AFE-PUND method is shown. It can be clearly understand that this curve has no physical meaning. 

\begin{figure}[htpb]
    \centering
    \includegraphics[width=0.47\textwidth]{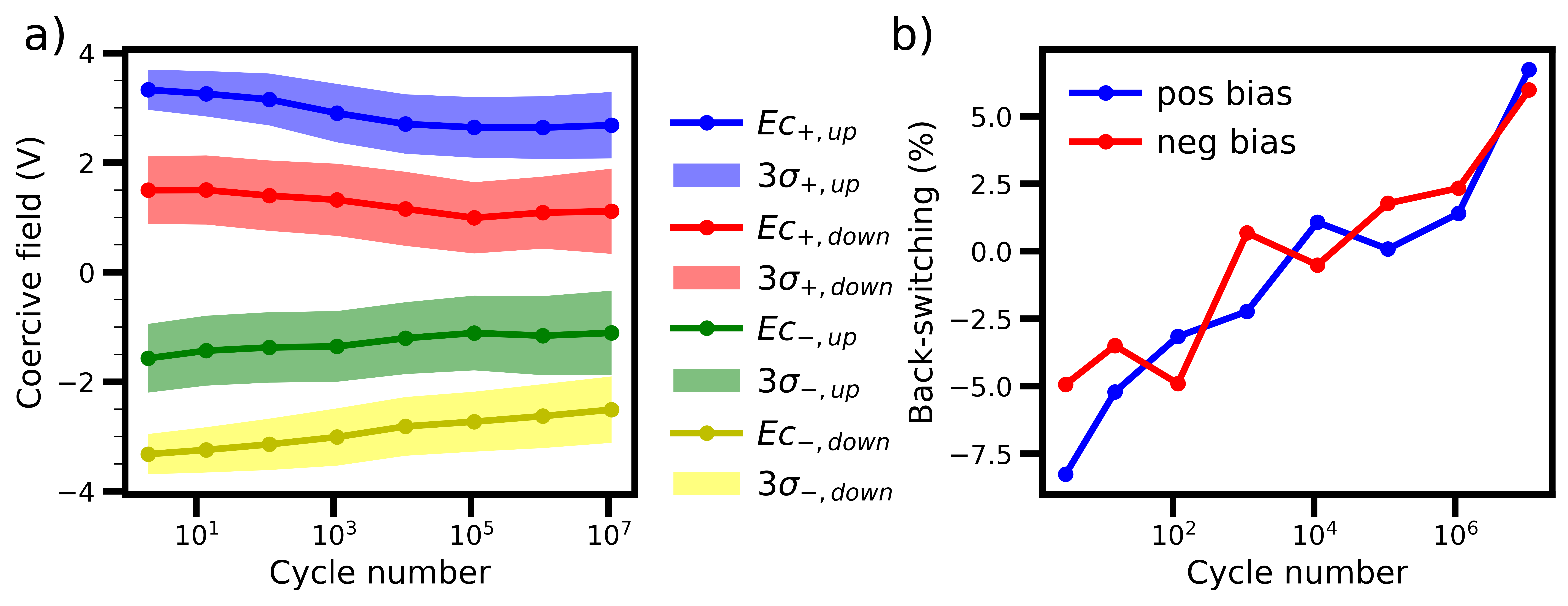}
    \caption{Back-switching limitations of the AFE-PUND method: (a) Coercive fields and switching peaks 3$\sigma$ (99.7\% of the switching) and (b) Back-Switching, as a function of cycle number on a $TiN$/$ZrO_2$/$TiN$ with 10 nm-thick $ZrO_2$ at 3.5 MV/cm on the positive applied bias side (blue) and negative side (red).}
    \label{fig:5}
\end{figure}

\indent 
    In a first approximation, when leakage is small enough, switching peak can be modelled with a Gaussian distribution:

\begin{equation}
    f(x) = \frac{1}{\sigma \sqrt{2 \pi}}e^{\frac{1}{2}\left(\frac{x-\mu}{\sigma}\right)^2}
\end{equation}    
    
\indent
    From the Gaussian model, we can extract $\sigma$ values. Switching dispersion of ferroelectric grains can be quantified by studying switching current at $3\sigma$ value. It represents the value for which 99.7 \% of the polarization is switched. On figure \ref{fig:5}(a) we can observe that the switching distribution around the coercive fields tends to increase with cycling. The structure also tends to symmetrize as $|E_{c,+}|$ and $|E_{c,-}|$ values tend to be closer and closer with cycling.
    \\
\indent
    The BS can be quantified as a function of the endurance if it is normalized using $P_r$: 

\begin{equation}
    BS = \frac{a}{P_r}\times 100
\end{equation}

    with BS expressed in percentage and a being the ratio between the polarization at rest: $V_{relax,+}$ and $V_{relax,-}$, and respectively the minimum and maximum polarization $P_{min}$ and $P_{max}$. For BS phenomenon, $P_{min}$ is negative when $V > 0$ and $P_{max}$ is positive when $V < 0$ while $P_r$ have an opposite sign. In other words: 

\begin{equation}
    a = \frac{P(V_{relax})}{P_{min,max}}
\end{equation}

\indent
    For more details and clarity, the calculations used to analyze BS mechanisms are illustrated more thoroughly in the supplemental. 
\\
\indent
    The miss-connection observed in the AFE-PUND curves, as observed in Figure \ref{fig:1}(f) and \ref{fig:3}(d) at the $V_{relax,+}$ and $V_{relax,-}$, is a direct result of BS phenomena. Upon relaxation, some ferroelectric domains spontaneously switch back to their original state, causing a residual switching current during the U or D pulses. This behavior is inherent to the material and becomes more prominent as the distribution of switching currents ($3\sigma$) begins to overlap. This overlap is indicative of significant BS and leads to inaccuracies in isolating the true ferroelectric response through the AFE-PUND method. Despite its unavoidable nature in certain material systems, BS can be quantified to evaluate its impact, and PUND methods (AFE and FE) ease its evaluation. It can be noted that because of back-switching a miss-connection is also observed for the PUND method applied to simple ferroelectrics as well. As shown in Figure \ref{fig:5}(b), the percentage of BS can be calculated, providing critical insights into the extent of this phenomenon and its influence on the extracted polarization values. This result is the same for antiferroelectrics and ferroelectrics. 

\section{Conclusion}

\indent
    In this article, we have presented an automatic pulse sequence  allowing to subtract switching current contributions with non-switching current contributions for antiferroelectric materials. $ZrO_2$ thin films of varying thicknesses were used as a practical example to demonstrate the effectiveness of the AFE-PUND method.
\\
\indent 
    Despite its robustness, the method has certain limitations, such as its sensitivity to high leakage currents and back-switching (BS). These challenges are not unique to AFE materials but are inherent to the PUND protocol itself. Addressing these issues, for example by developing intermediate sequences mimicking the Dynamic Leakage Current Compensation (DLCC) principle in order to compensate the leakage current during the measurement, will be an important focus of future work. 
\\
\indent 
    The current work provides a crucial framework for advancing the study of AFE materials. Additionally, exploring nanoscale measurements, similar to the nano-PUND protocol developed by Martin et al. \cite{martin2017new}, could further enhance our understanding by enabling high-resolution characterization of local ferroelectric and antiferroelectric behaviors using scanning probe microscopy techniques. Such developments will be vital for optimizing $ZrO_2$-based nano-supercapacitors and broadening their application in next-generation energy storage devices.

\bibliography{biblio.bib} 

\end{document}